\documentclass[12pt]{article}
\usepackage{lscape}
\usepackage{graphicx,epsfig}
\usepackage[tbtags]{amsmath}

\newcommand{\mathsc}[1]{\mathsf{\scriptstyle #1}}
\begin{document}

  \title{Nondecoupling Effects of Heavy Higgs Particles in Two Higgs Doublet Model
  \footnote{talk at the international conference RG2002, to be published soon 
  in Acta Physica Slovaca}}

\author{Michal Malinsk\'{y}\thanks{
Institute of Particle and Nuclear Physics, Charles University, Prague \newline
\mbox{}\hspace{5mm} E-mail: {\tt malinsky@ipnp.troja.mff.cuni.cz}
}}

\maketitle

  \abstract{Non-decoupling properties of additional heavy degrees of freedom in 
  the Higgs sector of the Two-Higgs-doublet 
  extension of the Standard model are discussed in a particular case of production 
  of a pair of longitudinaly polarized $W$-bosons in the $e^+e^-$ annihilation. 
  }



\section{Introduction}\label{sec:form}
\setcounter{section}{1}\setcounter{equation}{0}
\noindent

One of the least understood features of the Standard Model (SM) of electroweak 
interactions is the mechanism of electroweak symmetry breaking 
descending from the structure of the Higgs sector. 
Since the measurable quantities usually depend 
only weakly on its particular realization, many alternative models were proposed. 

Perhaps the most popular (nonsupersymmetric) extension of the SM Higgs sector is 
the so called Two Higgs doublet model (THDM, 2HDM) \cite{guide}, a theory with two 
Higgs doublets instead of one in the usual case. Not only it is capable to 
reproduce all the predictions of the standard theory but it also provides a nice framework 
for some possible new phenomena, among other things the CP-violation in the Higgs sector.   


Having two doublets in the model the number of Higgs degrees of freedom gets 
doubled, i.e. we are left with 5 (8 total - 3 Goldstone modes) physical Higgs states.
As in the case of SM the mass of the lightest state $h^0$ is expected to be at the 
electroweak scale but the other states $H^0$, $A^0$, $H^\pm$ can in principle be  
quite massive.
Although they cannot be produced in low-energy experiments 
it is legitimate to ask whether they could contribute to low-energy amplitudes by
means of virtual effects i.e. if they {\it decouple} in the heavy-mass limit or not.

There is a famous theorem by Appelquist and Carazzone \cite{appel} concerning
general decoupling properties of heavy degrees of freedom in field theories. 
Unfortunately it does not work well in the case of the couplings among the 
light and heavy sectors growing too fast with the heavy-sector masses. 
Note that the Higgs 
couplings are typically proportional to the masses of interacting particles 
and this spoils the validity of this theorem in many situations involving 
heavy virtual Higgses coupled to light sector in the game. 
This is not in general the case of SUSY theories in which the form
of the Higgs potential is rather strictly dictated by supersymmetry; 
from this point of view the 
heavy  Higgs particles in these theories decouple in a usual 
Appelquist-Carazzone manner, see \cite{dobado}

The problem of possibly large non-decoupling effects of heavy Higgs particles in 
THDM was discussed in several papers in 1990s, for instance \cite{haber} and 
\cite{kanem}.

As was shown in \cite{kanem} 
we can expect relatively large non-decoupling effects 
of heavy Higgs bosons in this model for instance in cross-sections of processes 
involving
longitudinal gauge bosons, in particular in
$e^+e^-\to W_L^+W_L^-$. The magnitude of deviation of this quantity compared to the 
well-known SM value turns out to be of the order of several percent, which may 
(at least in principle) be measurable at future facilities. 

In the calculation \cite{kanem} several simplifications have been made:
\\ 1. The 'Equivalence theorem' \cite{equiv} used therein works well 
only in the high-energy limit and therefore should not be used to estimate the 
non-decoupling features of the model (defined in the low-energy regime).
\\ 2. Only the ratio of total cross-sections is given
which effectively washes out all the interesting (and probably larger) 
effects in differential quantities.

From this point of view 
we find it meaningful to {\it recalculate the 
ratio of the differential cross-sections of 
$e^+e^-\to W_L^+W_L^-$ between THDM and SM} 
without use of the Equivalence theorem.

\section{General analysis}
Let us define the central quantity of our interest -- the ratio of the differential 
cross-sections of $e^+e^-\to W_L^+W_L^-$ in THDM and SM respectively:
\begin{equation} 
\delta \equiv \frac{{\rm 
d}\sigma^{\mathsc{thdm}}(e^+e^- \to W^+W^-)}{{\rm d}\sigma^{\mathsc {sm}}(e^+e^- 
\to W^+W^-)} - 1 
\end{equation}
Expanding now the THDM amplitude around the well-known SM value it is easy to 
obtain (at one-loop level)
\begin{equation}
\label{firststage}
\delta=2{\rm Re}
\frac{\Delta{\cal M}_{\mathsc{tree}}+\Delta{\cal M}_{\mathsc{1-loop}}}
{{\cal M}_{\mathsc{tree}}^{\mathsc{sm}}}
+\frac{k_2}{k_1}
\int {\rm d}k_\gamma
\frac                                       
{\left|{\cal B}^{\mathsc{thdm}}\right|^2-\left|{\cal B}^{\mathsc{sm}}\right|^2}
{\left|{\cal M}_{\mathsc{tree}}^{\mathsc{sm}} \right|^2}
 +
\ldots 
\end{equation}
Here the symbol ``$\Delta$'' denotes differences of given quantities between 
THDM and SM, for example  $\Delta{\cal M}_{\mathsc{1-loop}}$ is the difference of 
all one-loop contributions to amplitudes between the models; 
${\cal B}^{\mathsc{model}}$ denote the corresponding bremsstrahlung amplitudes 
needed to regulate the IR divergences of $\Delta{\cal M}_{\mathsc{1-loop}}$ 
and $k_i$'s are 
some geometrical factors.

First we can get rid of the bremsstrahlung part of this expression: the 
IR divergent parts of $\Delta{\cal M}_{\mathsc{1-loop}}$ are combined with differences 
of bremsstrahlung terms to give a perfectly finite quantity 
which is suppressed by a factor of $m_e/m_{\mathsc{w}}$ 
(Yukawa couplings of Higgses to electrons in the initial state) in comparison 
with the rest of (\ref{firststage}). Next, similar argumentation shows that the same 
proportionality factor occurs also in $\Delta{\cal M}_{\mathsc{tree}}$. Neglecting
such terms we are left with 
\begin{equation}
\label{basicrelation}
\delta \doteq 2{\rm Re}
\frac{
\Delta{\cal M}^{\mathsc{ir-fin.}}_{\mathsc{1-loop}}
}
{
{\cal M}_{\mathsc{tree}}^{\mathsc{sm}}
}
\end{equation}
All we need are therefore contributions of IR-finite one-loop graphs which are not common 
to both models.

\section{Two Higgs doublet model (THDM)}
Let us now specify the basic features of THDM in more detail. As we already know 
the presence of the second doublet gives rise to 5 Higgs states in the spectrum:
neutral scalars $h^0$ and $H^0$, charged scalars $H^\pm$ and a neutral 
pseudoscalar $A^0$. The most general form of the Higgs potential 
(in terms of $SU(2)$ doublets $\Phi_1$, $\Phi_2$, \cite{guide})
\begin{multline}
  V(\Phi_1,\Phi_2)= m_{11}^2\Phi_1^\dagger\Phi_1+m_{22}^2\Phi_2^\dagger\Phi_2-
		m_{12}^2\Phi_1^\dagger\Phi_2-m_{12}^{2*}\Phi_2^\dagger\Phi_1+  \\
		 +
		\frac{\lambda_1}{2}(\Phi_1^\dagger\Phi_1)^2+
		\frac{\lambda_2}{2}(\Phi_2^\dagger\Phi_2)^2+
		\lambda_3(\Phi_1^\dagger\Phi_1)(\Phi_2^\dagger\Phi_2)+ \\ +\lambda_4(\Phi_1^\dagger\Phi_2)(\Phi_2^\dagger\Phi_1)+ 
		\frac{\lambda_5}{2}(\Phi_1^\dagger\Phi_2)^2+
		\frac{\lambda_5^*}{2}(\Phi_2^\dagger\Phi_1)^2+\nonumber \\
		 +
		[\lambda_6(\Phi_1^\dagger\Phi_1)+\lambda_7(\Phi_2^\dagger\Phi_2)]
		(\Phi_1^\dagger\Phi_2)+ 
		[\lambda_6^*(\Phi_1^\dagger\Phi_1)+\lambda_7^*(\Phi_2^\dagger\Phi_2)]
		(\Phi_2^\dagger\Phi_1)\nonumber
\end{multline}
is less restrictive than in SUSY theories; there is enough freedom for
the coupling constants 
$\lambda_i$ and the 'mass-parameters' $m_{ij}$ to give rise to some new 
phenomena like the above mentioned CP-violation in the Higgs sector etc. 
Moreover, the bounds on the Higgs-mass patern are not so stringent  
as for example in the SUSY theories, it is not a problem to have 
the masses of $A^0$ or $H^\pm$ at a scale of several TeV \cite{guide},\cite{krawczyk}.


In general there are two basic types of THDM concerning the mass generation of up- and 
down-types of fermions. In type-I models both the up and down fermion masses are generated
by one of the doublets only in analogy with the minimal Higgs model while in type-II
theories one of the doublets generates the up-type and the second one the down-type masses 
in a similar way as in the MSSM. However, our analysis turns out 
to be model-independent because there are no relevant Yukawa couplings in the 
one-loop leading term computation. 

\section{Calculation of $\Delta{\cal M}^{\mathsc{ir-fin.}}_{\mathsc{1-loop}}$}

Although the full set of Feynman diagrams contributing to 
$\Delta{\cal M}^{\mathsc{ir-fin.}}_{\mathsc{1-loop}}$ is quite large, many of them
can be safely neglected. It is mainly due to the 
presence of a suppressing Yukawa factor in all the diagrams involving  Higgs couplings 
to electrons as indicated above.

Since we are using the on-shell renormalization scheme there is no need for 
renormalization of external legs. However, the vertex and propagator counterterms 
become nontrivial being not only the ``$1/\varepsilon$'' parts of the 
dimensionally-regularised expressions but fixed by on-shell renormalization 
conditions \cite{aoki}. These additional structures need their own detailed 
discussion. 

With these observations in mind we can classify all the one-loop topologies 
contributing to 
$\Delta{\cal M}^{\mathsc{ir-fin.}}_{\mathsc{1-loop}}$: it turns out that
the only really important 
types of graphs are the following:
   \begin{equation}
     \label{graphs}
       \epsfig{file=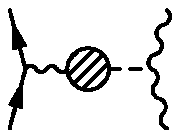,width=2cm}\hskip 1cm
       \epsfig{file=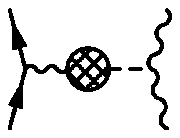,width=2cm}\hskip 1cm
       \epsfig{file=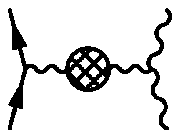,width=2cm}\hskip 1cm
       \epsfig{file=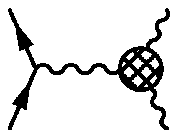,width=2cm}
   \end{equation}
Here the dark shaded blob corresponds to loops involving at least one Higgs 
boson while the 
lighter stands for loops without any Higgs inside, all of them including finite 
parts
of the relevant counterterms.

\subsection{Oblique-type corrections} \label{subsec:revi}
Although there is no direct suppression by the Yukawa factors 
in first two graphs in (\ref{graphs}) we can neglect them because of the one-loop 
mixed propagators whose contributions in general look like  
$\Pi(p^2,m_i^2)(p_1+p_2)_\alpha$.
Contracting this expression with the leptonic current and using the Dirac 
equation one reproduces again suppressing factor $m_e/m_{\mathsc{w}}$.
The case of the third topology is not so clear but it can be shown to exhibit the 
decoupling properties in the heavy (physical) Higgs mass limit \cite{kanem}. 
(This can be easily seen in the particular case of the 
on-shell renormalization scheme \cite{mal2}.)
				     
\subsection{Vertex corrections -- one-loop TGV differences}
Thus we are left with only the fourth topology in (\ref{graphs}). To proceed, 
we need the differences of one-loop renormalized triple gauge vertex structures
$\gamma W^+W^-$ and $ZW^+W^-$ \cite{mal1}; let us denote them by 
$\Delta \Gamma^\gamma_{\sigma\mu\nu}$  and 
$\Delta \Gamma^{\mathsc{Z}}_{\sigma\mu\nu}$ respectively. 
Relevant graphs can be divided into several clusters 
(charged bosons are denoted by a generic symbol $A^\pm$ 
(for example $G^\pm $ stands for the charged Goldtone bosons in $R_\xi$ gauge), while the neutral ones by $B$,$C$):
   \begin{equation}
     \label{graphs2}
       \epsfig{file=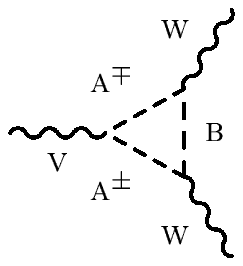,width=2cm}\hskip 1.5cm
       \epsfig{file=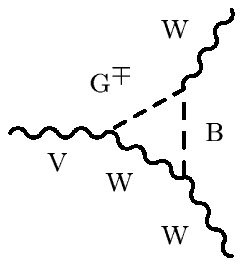,width=2cm}\hskip 1.5cm
       \epsfig{file=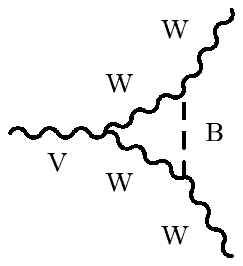,width=2cm}\hskip 1.5cm
   \end{equation}
\begin{displaymath} 
       \epsfig{file=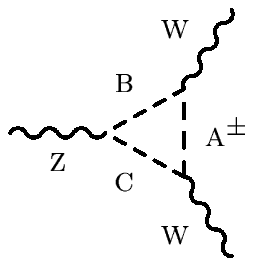,width=2cm}\hskip 1.5cm
       \epsfig{file=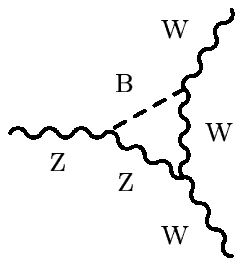,width=2cm}\hskip 1.5cm
       \epsfig{file=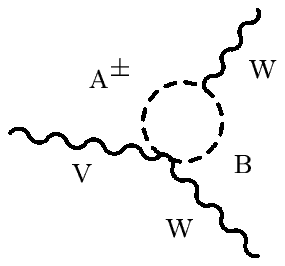,width=2cm}
\end{displaymath}
Their UV-divergences should be cancelled by counterterms descending from Ward-identity
$\delta Z_{\mathsc{w}}=\delta Z_g$ connecting the $W$-boson wavefunction renormalization constant 
 and the gauge coupling renormalization constant in the 
scheme fixed as in \cite{pokorski}.

Having everything at hand we can write the leading contribution to 
$\Delta{\cal M}^{\mathsc{ir-fin.}}_{\mathsc{1-loop}}$ in the form
\begin{equation}
\Delta{\cal M}^{\mathsc{ir-fin.}}_{\mathsc{1-loop}}\doteq 
\sum_{\mathsc{V}=\gamma,\mathsc{Z}} \bar{v}(p_1)\gamma_\lambda 
u(p_2)g_{ee\mathsc{v}}\frac{-ig^{\lambda\sigma}}{s-m^2_{\mathsc{V}}}
g_{\mathsc{vww}}\Delta\Gamma^{\mathsc{V}}_{\sigma\mu\nu}\,
\varepsilon^{*\mu}(q_1)\varepsilon^{*\nu}(q_2)
\end{equation}

\section{Results and conclusions}
Since we are dealing with many complicated diagrams 
(there are 46 graphs in an $R_{\xi=1}$ gauge in (\ref{graphs2}) and 9 others to calculate the finite 
parts of on-shell counterterms), we are forced to use a computer.
We have utilized {\tt Mathematica 4.0} with {\tt Feyncalc} and {\tt Looptools}. 
The figures Fig.1 and Fig.2 correspond to ratios of differential cross-sections of 
particular initial state helicity configurations $e_L^+e_R^- \to W_L^+W_L^-$
(in which $\delta$ is expected to be largest). Moreover, in this particular case the 
leading contribution to $\delta$ turns out to be $\cos \theta^*$-independent  
(CMS scattering angle).
Although we use a slightly different Higgs mass pattern to exhibit 
mainly the basic features of  $\delta$,  
its magnitude is in rough agreement with the expectation of \cite{kanem}.
(The discontinuities in derivatives of the second curve originate 
from the fact, that the loop integrals in (\ref{basicrelation}) acquire non-zero 
imaginary parts above some values of $\sqrt{s}$ which correspond to tresholds
of productions of the loop particles in on-shell final states.)

\begin{figure}[h]
\begin{center}   
        \epsfig{file=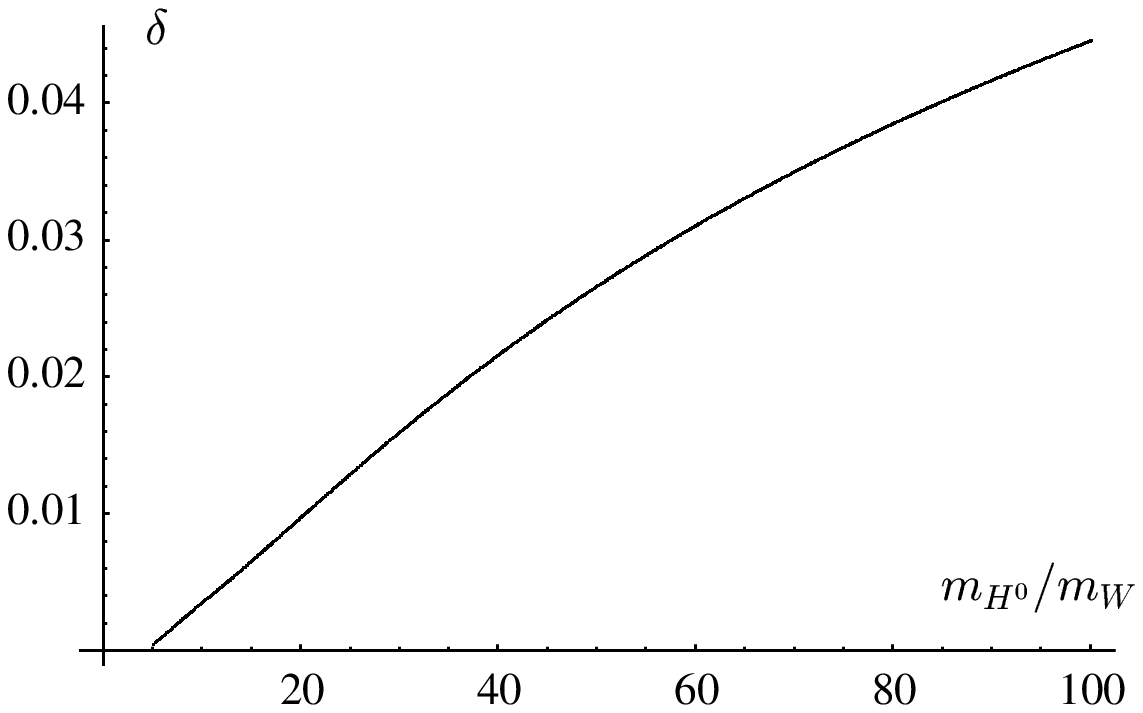,width=6cm,height=3cm}
\\ 
\caption{this plot shows $\delta[e_L^+e_R^- \to W_L^+W_L^-]$ 
as a function of $m_{H^0}/m_{\mathsc{w}}$ \newline
Other parameters:  
${\sqrt{s}\doteq}$ 650GeV, 
$m_{h^0}\sim m_{\eta}\doteq$ 130GeV, 
${m_{A^0}\doteq}$ 4TeV, 
${m_{H^\pm}\doteq}$ 2TeV 
}
\end{center}
\end{figure}
\begin{figure}[h]
\begin{center}   
        \epsfig{file=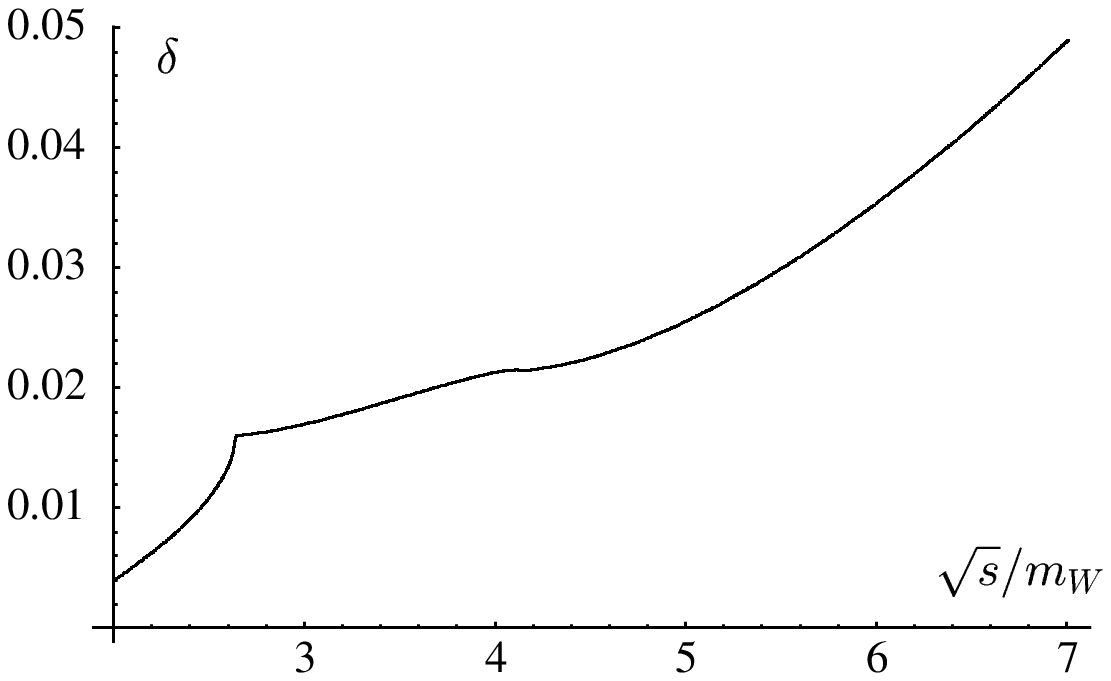,width=6cm,height=3cm}
\\ \caption{this plot shows 
$\delta[e_L^+e_R^- \to W_L^+W_L^-]$ 
as a function of $\sqrt{s}/m_{\mathsc{w}}$ \newline
Other parameters:  
$m_{h^0}\sim m_{\eta}\doteq$ 130GeV, 
$m_{H^0}\doteq$ 240GeV, 
${m_{A^0}\doteq}$ 8TeV, 
${m_{H^\pm}\doteq}$ 800GeV 
}
\end{center}
\end{figure}
As we see, at least in some regions of the parametric space we can expect 
relatively large non-decoupling (note the slope of the plot at Fig.1) 
effects of heavy Higgs bosons in the 
considered quantity; they can reach the order of several percent. 
\\
\vskip 2mm
{\bf Acknowledgement}
\\
I would like to thank to  Prof. Ji\v{r}\'{\i} Ho\v{r}ej\v{s}\'{i} 
for all his comments and helpful suggestions during the preparation of
this contribution. This work was supported by ``Centre for Particle Physics'', 
project No. LN00A006 of the Ministry of Education of the Czech Republic.

\end{document}